\documentclass{article}
\usepackage{geometry} 
\usepackage{amsmath}
\usepackage[latin1]{inputenc}
\usepackage[OT1]{fontenc}

\newtheorem{definition}{Definition}
\newtheorem{theorem}{Theorem}
\newtheorem{example}{Example}

\newcommand{\mathconstante}[1]{\ensuremath{{\mathrm{#1}}}}
\newcommand{\Constants}{\ensuremath{C}}
\newcommand{\Variables}{\ensuremath{\call{X}}}
\newcommand{\fonction}[2]{\ensuremath{{\mathconstante{#1}(#2)}}}
\newcommand{\termset}[1]{{\fonction{T}{#1}}} 
\newcommand{\gsig}[1]{\termset{#1}}
\newcommand{\vsig}[1]{\termset{#1,\Variables}}
\newcommand{\call}[1]{\ensuremath{\mathcal{#1}}}

\newcommand{\nonces}[1]{\fonction{nonces}{#1}}
\newcommand{\rinput}[1]{\fonction{input}{#1}}
\newcommand{\strand}[1]{\fonction{strand}{#1}}
\newcommand{\TFX}{\vsig{\call{F}}}

\newcommand{\set}[1]{\ensuremath{{\left\lbrace #1 \right\rbrace}}}
\newcommand{\Var}[1]{\fonction{Var}{#1}}

\newcommand{\Cons}[1]{\fonction{Cons}{#1}}
\newcommand{\Supp}[1]{\fonction{Supp}{#1}}
\newcommand{\tq}{\ensuremath{\,|\,}}
\newcommand{\condset}[2]{\set{#1\tq{}#2}}

\newcommand{\unif}{\ensuremath{\stackrel{?}{=}}}

\newcommand{\penc}[2]{\ensuremath{ \text{\rm enc}(#1,#2)}}
\newcommand{\pdec}[2]{\ensuremath{\text{\rm dec}( #1,#2)}}
\newcommand{\paire}[2]{\ensuremath{\left\langle #1,#2\right\rangle}}
\newcommand{\piun}[1]{\ensuremath{\pi_1(#1)}}
\newcommand{\pideux}[1]{\ensuremath{\pi_2(#1)}}
\newcommand{\msg}[2]{\ensuremath{\text{\rm msg}( #1,#2)}}
\newcommand{\payload}[1]{\ensuremath{\text{\rm payload}(#1)}}
\newcommand{\partner}[1]{\ensuremath{\text{\rm partner}(#1)}}
\newcommand{\symtest}[2]{\ensuremath{\mathrm{symtest}({#1},{#2})}}
\newcommand{\pairtest}[1]{\ensuremath{\mathrm{pairtest}({#1})}}
\newcommand{\lmodel}{[\![}
\newcommand{\rmodel}{]\!]}
\newcommand{\notark}[1]{}

\title{Compiling and securing cryptographic protocols}
\date{\today}
\author{Yannick Chevalier, Micha{\"e}l Rusinowitch\\
\small Loria, Inria Nancy Grand Est\\[-0.8ex]
\small Campus Scientifique --- BP 239 54506 Vand\oe uvre-l{\`e}s-Nancy, France\\
\small \texttt{\{chevalie,rusi\}@loria.fr}\\
}

\begin{document}
\bibliographystyle{elsarticle-num}
\sloppy

\maketitle


  \begin{abstract}
    Protocol narrations are widely used in security as semi-formal
    notations to specify conversations between roles.  We define a
    translation from a protocol narration to the sequences of
    operations to be performed by each role.  Unlike previous works,
    we reduce this compilation process to well-known decision problems
    in formal protocol analysis. This allows one to define a natural
    notion of prudent translation and to reuse many known results from
    the literature in order to cover more crypto-primitives. In
    particular this work is the first one to show how to compile
    protocols parameterised by the properties of the available
    operations.
  \end{abstract}

\section{Introduction}

\begin{figure}[htbp]
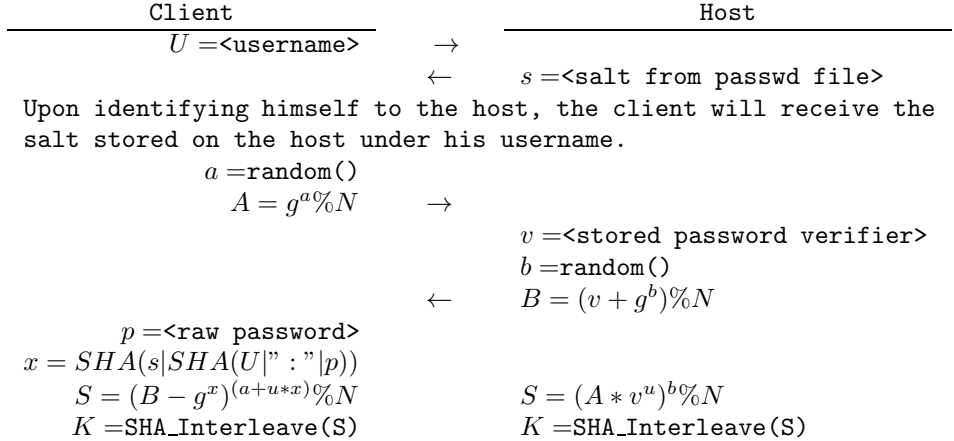

  \centering{\tt
  \begin{tabular}{rcl}
  \multicolumn 1c {Client} & & \multicolumn 1c  {Host}\\
  \cline{1-1}\cline{3-3}
  $U = $<username>          &~~~$\rightarrow$~~~&\\
                          &$\leftarrow$&    $s = $<salt from passwd file>\\
\multicolumn{3}{l}{Upon identifying himself to the host, the client will receive the}\\
\multicolumn{3}{l}{salt stored on the host under his username.}\\
      $a = $random() &&\\
      $A = g^a \% N$          &$\rightarrow$&\\
                     &&                    $v = $<stored password verifier>\\
                     &&                    $b = $random()\\
                                  &$\leftarrow$&    $B = (v + g^b) \% N$\\
      $p = $<raw password>&&\\
      $x = SHA(s | SHA(U | ":" | p))$ &&\\
      $S = (B - g^x) ^ {(a + u * x)} \% N$  &&  $S = (A * v^u) ^ b \% N$\\
      $K = $SHA\_Interleave(S)            &&  $K = $SHA\_Interleave(S)\\
    \end{tabular}}
  \caption{\label{fig:srp}Annotated message sequence chart extracted from the RFC~2945 (SRP Authentication and Key Exchange System)}
  
\end{figure}

Cryptographic protocols are designed to prescribe message exchanges
between agents in hostile environment in order to guarantee some
security properties such as confidentiality. There are many apparently
similar ways to describe a given security protocol. However one has to
be precise when specifying how a message should be interpreted and
processed by an agent since overlooking subtle details may lead to
dramatic flaws.  The main issues are the following:
\begin{itemize}
\item What parts of a received message should be extracted and checked
  by an agent?
\item What actions should be performed by an agent to compute an
  answer?
\end{itemize}
These questions are often either partially or not at all adressed in
common protocol descriptions such as the so-called protocol
narrations.  A \emph{protocol narration} is the definition of a
cryptographic protocol by the intended sequence of messages. For
example the well-known Needham-Schroeder Public Key
protocol~\cite{nspk78} is conveniently specified by the following
text:
$$
\begin{array}{c@{\rightarrow}c@{:}l}
  A & B & \penc{\paire A {N_a} }{K_B} \\
  B & A & \penc{\paire  {N_a} {N_b} }{K_A} \\
  A & B & \penc{ N_b }{ K_B } \\
  \multicolumn 3l {\text{\bf where }}\\
  \multicolumn 3l {A\textbf{ knows }A,B,K_A,K_B,K_A^{-1}} \\
  \multicolumn 3l {B\textbf{ knows }A,B,K_A,K_B,K_B^{-1}} \\
\end{array}
$$
Protocol narrations are also a textual representation of Message
Sequence Charts (MSC), which are employed \textit{e.g.} in RFCs. For
more complex protocols, one needs to indicate the internal
computations of each participant either by annotating the MSC or by
employing the Lowe operator~\cite{lowe-operator} or otherwise express
internal actions that have to be performed, as in the specification of
Fig.~\ref{fig:srp}.

We claim that all internal computations specified in
Figure~\ref{fig:srp}, and more generally most such annotations, can be
computed automatically from the protocol narration.  Our goal in this
paper is to give an operational semantics to---or, equivalently, to
\emph{compile}---protocol narrations so that internal actions
(excluding \textit{e.g.}  storing a value in a special list for a use
external to the protocol) are described.


\paragraph{Related works} 
Although many works have been dedicated to verifying cryptographic
protocols in various formalisms, only a few have considered the
different problems of extracting operational (non ambiguous) role
definitions from protocol descriptions.  Operational roles are
expressed as multiset rewrite rules in CAPSL~\cite{millen02},
CASRUL~\cite{jrv00}, or sequential processes of the spi-calculus with
pattern-matching~\cite{cvb-tcs06}.  This extraction is also used for
end-point projection~\cite{McCarthy07,McCarthy08}.  A pioneering work
in this area is one by Carlsen~\cite{Carlsen} that has proposed a
system for translating protocol narrations into
CKT5~\cite{Bieber-csfw}, a modal logic of communication, knowledge and
time.

Compiling narrations to roles has been extended beyond perfect
encryption primitives to algebraic theories in
\cite{yannickthesis,zeb}.  We can note that, although these works
admit very similar goals, all their operational role computations are
ad-hoc and lack of a uniform principle. In particular they essentially
re-implemented previously known techniques. An advantage of \cite{zeb}
is that it supports implicit decryption which may lead to more
efficient secrecy decision procedures.

We propose here a uniform approach to role computation that allows us
to relate the problem to well-known decision results in formal
cryptographic protocols analysis, namely the \emph{reachability
  problem}. Moreover this approach is also used successfully for the
automatic computation of prudent security wrapper (a.k.a. security
tests) for filtering messages received by principals.  We show how to
reduce this computation to known results about the standard notion of
\emph{static equivalence}.

\section{Role-based Protocol Specifications}
\label{sec:role}

First we show how from Alice\& Bob notation we can derive a plain role-based
specification. Then the specification will be refined 
in the following Sections. 

\subsection{Specification of messages and basic operations}

\paragraph{Terms}
We consider an infinite set of free constants \Constants{} and an
infinite set of variables \Variables. For each signature \call{F}
(\textit{i.e.} a set of function symbols with arities), we denote by
\gsig{\call{F}} (resp.  \vsig{\call{F}}) the set of terms over
$\call{F}\cup{}\Constants{}$ (resp.
$\call{F}\cup{}\Constants{}\cup\Variables$).  The former is
called the set of ground terms over \call{F}, while the later is
simply called the set of terms over \call{F}.  Variables are denoted
by $x$, $y$, terms are denoted by $s$, $t$, $u$, $v$, and finite sets
of terms are written $E,F,...$, and decorations thereof, respectively.
We abbreviate $E\cup F$ by $E,F$, the union $E\cup\set{t}$ by $E,t$
and $E\setminus \set{t}$ by $E\setminus t$.

In a signature \call{F} a \emph{constant} is either a free constant or a
function symbol of arity $0$ in \call{F}. Given a term $t$ we denote by \Var{t}
the set of variables occurring in $t$ and by \Cons{t} the set of constants
occurring in $t$.  A substitution $\sigma$ is an idempotent mapping from
\Variables{} to \vsig{\call{F}} such that $\Supp{\sigma}=
\condset{x}{\sigma(x)\not=x}$, the \emph{support} of $\sigma$, is a finite set.
The application of a substitution $\sigma$ to a term $t$ (resp.  a set of terms
$E$) is denoted $t\sigma$ (resp.  $E\sigma$) and is equal to the term $t$
(resp. $E$) where all variables $x$ have been replaced by the term $x\sigma$. A
substitution $\sigma $ is \emph{ground}  if for each $x\in
\Supp{\sigma }$ we have $x\sigma \in \gsig{\call{F}}$. 

\textsl{Operations.} Terms are manipulated by applying
\emph{operations} on them. These operations are defined by a subset of
the signature \call{F} called the \emph{set of public constructors}. A
context $C[x_1,\ldots,x_n]$ is a term in which all symbols are public
and such that its  nullary symbols   are either public non-free constants or
variables.
 
\textsl{Equational theories.} An \emph{equational presentation}
$\call{E}=(\call{F},E)$ is defined by a set $E$ of equations $u=v$
with $u,v\in\vsig{\call{F}}$. The equational theory generated by
$(\call{F},E)$ on $\vsig{\call{F}}$ is the smallest congruence
containing all instances of axioms of $E$ (free constants can also 
be used for building instances).  We write $s =_{\call{E}}t$
as the congruence relation between two terms $s$ and $t$.  By abuse of 
terminology we also call $\call{E}$ the equational theory generated by
the presentation \call{E} when there is no ambiguity. This equational
theory is introduced in order to specify the effects of operations on the
messages and the properties of messages.

\textsl{Deduction systems.}  A \emph{deduction system} is defined by a
triple $( \call{E},\call{F},\call{F}_p)$ where \call{E} is an
equational presentation on a signature \call{F} and $\call{F}_p$ a
subset of \emph{public} constructors in \call{F}.  For instance the
following deduction system models public key cryptography: $( \set{
  \pdec{\penc{x}{y}}{y^{-1}} = x } ,\set{ \pdec{\_}{\_}
  ,\penc{\_}{\_}, {\_^{-1}} },\set{\pdec{\_}{\_},\penc{\_}{\_} } )$
The equational theory is reduced here to a single equation that
expresses that one can decrypt a ciphertext when the inverse key is
available.

\subsection{Role Specification}

We present in this subsection how protocol narrations are transformed
into  sets of \emph{roles}. A role can be viewed as the projection of the protocol 
on a principal. The core of a role is a \emph{strand} which is a standard notion 
in cryptographic protocol modeling~\cite{strand}.

A \emph{strand} is a finite sequence of messages each with label (or polarity) $!$ or $?$.
Messages with label $!$ (resp. $?$) are said to be ``sent'' (resp.``received''). 
A strand is \emph{positive} iff all its labels  are $!$. 
Given a list of message $l= m_1, \ldots , m_n$ we write $?l$ (resp.  $!l$) as a short-hand 
for $?m_1, \ldots , ?m_n$, (resp. $!m_1, \ldots , !m_n$).

\begin{definition}{}\label{def:role}
  A \emph{role specification} is an expression
  $A(\vec{l}):\nu\vec{n}.(S)$ where $A$ is a name, $\vec{l}$ is a
  sequence of constants (called the \emph{role parameters}),
  $\vec{n}$ is a sequence of constants (called the \emph{nonces} of
  the role), and $S$ is a strand. Given a role $r$ we denote by
  $\nonces{r}$ the nonces $\vec{n}$ of $r$ and ~\strand{r} the strand
  $S$ of $r$.
\end{definition}

\begin{example}{\label{ex:role:specification}}
  For example, the initiator of the NSPK protocol is modeled, at this
  point, with the role:
  $$
  \begin{array}{l}
    \nu N_a.(? N_a,  ? A,? B,? K_A,? K_B,? K_A^{-1},\\
    ! \msg{B}{\penc{\paire A {N_a} }{K_B}}, ? \msg B {\penc{\paire {N_a} {N_b} }{K_A}},\\
    ! \msg B {\penc{ N_b }{ K_B }})
  \end{array}
  $$
  with the equational theory of public key cryptography, plus the
  equations \set{\piun{\paire xy} =x, \pideux{\paire xy} = y}.
\end{example}

Note that nothing guarantees in general that a protocol defined as a
set of roles is executable. For instance some analysis is necessary to
see whether a role can derive the required inverse keys for examining
the content of a received ciphertext. We also stress that role
specfications do not contain any variables. The symbols $N_a,A,\ldots$
in the above example are constants, and the messages occurring in the
role specification are all ground terms.

\paragraph{Plain roles extracted from a narration} From a protocol
narration where each nonce originates uniquely we can extract almost
directly a set of roles, called plain roles as follows. The constants
occurring in the initial knowledge of a role are the parameters of the
strand describing this role.  We model this initial knowledge by a
sequence of receptions (from an unspecified agent) of each term in the
initial knowledge.  In order to encode narrations we assume that we
have in the signature three public function symbols \msg\_\_,
\partner\_ and \payload\_ satisfying the equational theory:
$$
\left\lbrace
  \begin{array}{rcl}
    \partner{\msg xy} & = & x\\
    \payload{\msg xy} & = & y\\
  \end{array}
\right.
$$
For every agent name $A$ in the protocol narration, a role specification for $A$
is $A(l): \nu \nonces{S}.
(?\nonces{S},?K,S^A)$, where $K$ is such that $A\textbf{ knows }K$ occurs in the
protocol narration, $l$ is the set of constants in $K$. \nonces{S} and
strand $S^A$ are computed as follows:
\begin{description}
\item[Computation of $S^A$:]
  Init $S_0^A = \emptyset $\\
  On the $(n+1)$-th line $S\rightarrow R: M$ do
    $$
    S_{n+1}^A=\left\lbrace
      \begin{array}[c]{lr}
        S_n,!\msg{R}{M} & \text{If }A=S\\
        S_n,?\msg{S}{M} & \text{If }A=R\\
        S_n^A             & \text{Otherwise}\\
      \end{array}
    \right.
    $$

  \item[Computation of $\nonces{A}$:] This set contains each constant $N$ that
    appears in the strand $?K,S^A$ inside a message labelled $!$ and such that
    $N$ does not occur in previous messages (with any polarity).
\end{description}
This computation always extracts role specifications from a given protocol
narration and it has the property that every constant appears in a received
message before appearing in a sent message. Since a nonce is to be created
within an instance of a role, we reject protocol narrations from which the
algorithm described above extracts two different roles $A$ and $B$ with
$\nonces{A}\cap\nonces{B}\neq\emptyset$.

Example {\ref{ex:role:specification}} is a plain role that can be
derived by applying the algorithm to the NSPK protocol narration. We
now define the \emph{input} of a role specification which informally
is the sequence of messages \emph{sent to} a role as defined by the
protocol narration.


\begin{definition}{}
  Let $r=\nu N .(\frac !? M_i)_{1\le i\le n}$ be a role specification,
  and let $(R_1,\ldots,R_k)$ be the subsequence of the messages $M_i$
  labeled with $?$. 
The \emph{input} of $r$ is denoted \rinput{r} and is the
    positive strand $(!R_1,\ldots,!R_k)$.
\end{definition}

In the next section we define a target for the compilation of role
specifications.  Then we compute constraints to be satisfied by sent
and received messages.  and by adding the constraints to the
specification this one gets executable in the safest way as possible
w.r.t. to its initial specification.

\section{Operational semantics for roles}
In Section~\ref{sec:role}  we have defined roles and shown how 
they can be extracted from protocol narrations. 
In this section we define what  an implementation 
of a role is and in Section~\ref{sec:compil}  we will show how 
to compute such an implementation from a protocol narration.

\paragraph{Unification systems} 

Intuitively an operational model for
a role  has to reflect the possible manipulations on messages performed by a
program implementing the role. These operations  are specified here by 
a deduction system \call{D} $=(\call{E}, \call{F},\call{S})$   where 
the set of public functions  \call{S}, a subset of the 
signature  \call{F},  is defined by  equations in \call{E}. 
Beside defining function computations, the  equations \call{E}  specify some properties.

\begin{definition}{\label{def:unification}}
  Let $\call{E}$ be an equational theory.  An
  $\call{E}$-\emph{Unification system} $S$ is a finite set of
  equations denoted by $(u_i \unif{}
  v_i)_{i\in\set{1,\ldots,n}}$ with terms $u_i,v_i\in\TFX$.  It is
  satisfied by a substitution $\sigma$, and we note $\sigma\models{}
  S$, if for all $i\in\set{1,\ldots,n}$ $u_i\sigma =_\call{E}
  v_i\sigma$.
\end{definition}

\paragraph{Active frames} We introduce now the set of implementations of a role
specification as \emph{active frames}.  An active frame extends the role notion
by specifying how a message to be sent is constructed from already known
messages, and how a received message is checked to ascertain its conformity
w.r.t. already known messages. The notation $!v_i$ (resp. $?v_i$) refers to a
message stored in variable $v_i$ which is sent (resp. received).

\begin{definition}{\label{def:active:frame}}
  Given a deduction system \call{D} with equational theory \call{E}, a
  \call{D}-\emph{active frame} is a sequence $(T_i)_{1\le i \le k}$
  where
  $$
  T_i=\left\lbrace
    \begin{array}[c]{l@{\hspace*{-1pt}}r}
      !v_i \text{ with } v_i \unif{} C_i[v_1,\ldots,v_{i-1}] & \text{(send)}\\
      \multicolumn 2c {\text{\bf or}}\\
      ?v_i \text{ with } S_i(v_1,\ldots,v_{i}) & \text{(receive)}
    \end{array}
  \right.
  $$
  where $C_i[v_1,\ldots,v_{i-1}]$ denotes a context over variables
  $v_1,\ldots,v_{i-1}$ and $S_i(v_1,\ldots,v_{i})$ denotes a
  \call{E}-unification system over variables $v_1,\ldots,v_{i}$. 
  Each variable $v_i$ occuring with polarity $?$ is an \emph{input variable} of the active frame. 
\end{definition}

\begin{example}{\label{ex:active:frame}}
  The following is an active frame denoted $\phi_{a}$ that can be employed to
  model the role $A$ in the NSPK protocol:
$$
\begin{array}{l}
  (? v_{N_a} ? v_{A},? v_{B},? v_{K_A},? v_{K_B},? v_{K_A^{-1}},\\
  ! v_{msg_1} \text{ with } v_{msg_1}\unif \msg{v_{B}}{\penc{\paire {v_{A}} {v_{N_a}} }{v_{K_B}}},\\
  ? v_r \text{ with }  \emptyset \\ 
  ! v_{msg_2} \text{ with } v_{msg_2}\unif \msg {v_{B}} {\penc{
      \pideux{\pdec{v_r}{v_{K_A^{-1}}}} }{ v_{K_B} }})
\end{array}
$$
\end{example}

Compilation is the computation of an active frame from a role specification
such that, when receiving messages as intended by the role specification, the
active frame emits responses equal modulo the equational theory to the responses
issued in the role specification. More formally, we have the following:

\begin{definition}{\label{def:evaluation}}
  Let \call{D} be a deduction system with equational theory \call{E}.
  Let $\varphi=(T_i)_{1\le i \le k}$ be an active frame, where the
  $T_i$'s are as in Definition \ref{def:active:frame}, and where the
  input variables are $r_1,\ldots,r_n$. Let $s$ be a positive strand
  $!M_1,\ldots,!M_n$. Let $\sigma_{\varphi,s}$ be the substitution
  \set{r_i\mapsto M_i} and $S$ be the union of the
  \call{E}-unification systems in $\varphi$.  The \emph{evaluation} of
  $\varphi$ on $s$ is denoted $\varphi\cdot s$ and is the strand $(
  m_1,\ldots, m_k)$ where:
  $$
  m_i=\left\lbrace
    \begin{array}[c]{ll} 
      !C_i[m_1,\ldots,m_{i-1}] & \text{If }v_i\text{ has label ! in } T_i\\
     ?v_i\sigma_{\varphi,s} & \text{If }v_i\text{ has label ? in }  T_i\\
    \end{array}\right.
  $$
  We say that $\varphi$ \emph{accepts} $s$ if $S\sigma_{\varphi,s}$ is 
  satisfiable.
\end{definition}

To simplify notations, the application of a
\call{D}-context $C[x_1,\ldots,x_n]$ on a positive strand
$s=(!t_1,\ldots,!t_n)$ of length $n$ is denoted $C\cdot s$ and is the
term $C[t_1,\ldots,t_n]$. 
\begin{example}{\label{ex:frame:application}}
  Let $r$ be the role specification of role $A$ in NSPK as given in
  Ex.~\ref{ex:role:specification} and $\phi_A$ be the active frame of
  Ex.~\ref{ex:active:frame}. We have:
  $$
  \rinput{r}=(! N_a,  ! A,! B,! K_A,! K_B,! K_A^{-1},\\
     ! \msg B {\penc{\paire {N_a} {N_b} }{K_A}})
  $$
  and $\phi_A\cdot\rinput{r}$ is the strand:
  $$
  \begin{array}{l}
    (?N_a, ? A,?B,?K_A,?K_B,? K_A^{-1}, 
    !\msg{B}{\penc{\paire {A} {N_a} }{K_B}},
    ?\msg B {\penc{\paire {N_a} {N_b} }{K_A}}, \\
    !\msg {B} {\penc{ \pideux{\pdec{ \payload{\msg B {\penc{\paire {N_a} {N_b} }{K_A}} } }{K_A^{-1}}} }{ K_B }}
  \end{array} 
  $$
  Modulo the equational theory, this strand is equal to the strand:
  $$
  \begin{array}{l}
    (?N_a,  ?A,?B,?K_A,?K_B, ?K_A^{-1},
    !\msg{B}{\penc{\paire {A} {N_a} }{K_B}},
    ?\msg B {\penc{\paire {N_a} {N_b} }{K_A}},\\
    !\msg {B} {\penc{ N_b }{ K_B }}
  \end{array}
  $$
\end{example}
It is not coincidental that in Ex.~\ref{ex:frame:application} the
strands $\varphi\cdot\rinput{r}$ and $\strand{r}$ are equal as it
means that within the active frame, the sent messages are composed
from received ones in such a way that when receiving the messages
expected in the protocol narration, the role responds with the
messages intended by the protocol narration. This fact gives us a
criterion to define functional implementations of a role.

\begin{definition}{\label{def:implementation}}
  An active frame $\varphi$ is an \emph{implementation} of a role specification
  $r$ if $\varphi$ accepts $\rinput{r}$ and
  $\varphi\cdot\rinput{r}=_{\call{E}}\strand{r}$.  If a role admits an
  implementation we say this role is \emph{executable}.
\end{definition}

\paragraph{Example} $\phi_{a}$ defined above is a possible
implementation of the initiator role in NSPK.  However this
implementation does not check the conformity of the messages with the
intended patterns, \textit{e.g.} it neither checks that $v_r$ is
really an encryption with the public key $v_{K_A}$ of a pair, nor that
the first argument of the encrypted pair has the same value as the
nonce $v_{N_a}$. In Section~\ref{sec:compil} we show not only how to
compute an active frame when the role specification is executable, but
also to ensure that all the possible checks are performed.

\section{Compilation of role specifications}
\label{sec:compil}
Usually the compilation of a specification is defined by a
\emph{compilation algorithm}. An originality of this work is that we
present the result of the compilation as the solution to decision
problems. This has the advantage of providing for free a notion of prudent implementation as explained below.

\subsection{Computation of a ``vanilla''  implementation}

Let us first present how to compute an implementation of a role
specification in which no check is performed, as given in the
preceding example. To build such an implementation we need to 
compute for every sent message $m$ a context $C_m$ that evaluates to $m$ when  applied 
to the previously received ones. This reachability problem is unsolvable 
in general. Hence we have to consider  systems  that admit 
a reachability algorithm, formally defined below:

\begin{definition}{}
  Given a deduction system \call{D} with equational theory \call{E},
  a \emph{\call{D}-reachability algorithm} $\call{A}_{\call{D}}$ 
  computes, given a positive strand $s$ of length $n$ and a term $t$, a  \call{D}-context 
 $\call{A}_\call{D}(s,t) = C[x_1,\ldots,x_n]$ such that $C\cdot s=_{\mathcal{E}} t$ 
 iff there exists such a context and $\bot$ otherwise. 
\end{definition}

We will show that several interesting theories admit a reachability algorithm. 
This algorithm can be employed as an oracle to compute the contexts in
sent messages and therefore to derive an implementation of a role specification $r$. We
thus have the following theorem.
\begin{theorem}{\label{theo:executable}}
  If there exists a \call{D}-reachability algorithm 
  then it can be  decided whether a
  role specifications $r$ is executable and, if so  one can compute an implementation of $r$.
\end{theorem}

\paragraph{Proof sketch.} Let $r=(\frac !? M_i)_{i\in\set{1,\ldots,n}}$ be an
executable role specification. By definition there exists an active frame
$\varphi$ that implements $r$, \textit{i.e.} for each sent message $M_i$, there
exists a context $C_i$ such that $C_i[M_1,\ldots,M_{i-1}]$ is equal to $M_i$
modulo the equational theory. Thus if there exists a \call{D}-reachability
algorithm $\call{A}_\call{D}$, the result
$\call{A}_\call{D}(M_1,\ldots,M_{i-1}),M_i)$ cannot be $\bot$ by definition. As
a consequence, $\call{A}_\call{D}((M_1,\ldots,M_{i-1}),M_i)$ is a context
$C'_i[x_1,\ldots,x_n]$. Thus for all index $i$ such that $M_i$ is sent we can
compute a context $C'_i$ that, when applied on previous messages, yields the
message to send. We thus have an implementation of the role specification.

\subsection{Computation of a prudent implementation}

Computing an active frame is not enough since one would want to model
that received messages are checked as thoroughly as possible.  For
instance in Example~\ref{ex:role:specification}, a prudent
implementation of the message reception ``$? v_r \text{ with }
\emptyset$'' should be:
$$
? v_r \text{ with } \piun{\pdec{\payload{v_r}}{v_{K_A^{-1}}}} \unif
v_{N_a} \wedge \partner{v_r} \unif v_{B}
$$ 
Let us first formalize this by a refinement relation on sequences of
messages. We will say a strand $s$ refines a strand $s'$ if any
observable equality of subterms in strand $s$ can be observed in $s'$
using the same tests. To put it formally:

\begin{definition}{\label{def:refinement}}
  A positive strand $s=(!M_1,\ldots,!M_n)$ \emph{refines} a positive
  strand $s'=(!M_1',\ldots,!M_n')$ if, for any pair of contexts
  $(C_1[x_1,\ldots,x_n], C_2[x_1,\ldots,x_n])$ one has $ C_1\cdot s'
  =C_2\cdot s'$ implies $C_1\cdot s=C_2\cdot s$.
\end{definition}

For instance the strand $s=
(!\penc{\penc{a}{k'}}{k},!\penc{a}{k'},!k,!k',!a)$ refines $s'=
(!\penc{\penc{a}{k'}}{k},!\penc{a}{k'},!k,!k'',!a)$ since all
equalities that can be checked on $s'$ can be checked on $s$.  We can
now define an implementation to be \emph{prudent} if every equality
satisfied by the sequence of messages of the protocol specification is
satisfied by \emph{any} accepted sequence of messages.
\begin{definition}{\label{def:eq:prudent}}
  Let $r$ be a role specification and $\varphi$ be an implementation
  of $r$. We say that $\varphi$ is \emph{prudent} if any
  positive strand $s$ accepted by $\varphi$ is a refinement of
  \rinput{r}.
\end{definition}
As we shall see in Section~\ref{sec:appli}, most deduction systems
considered in the context of cryptographic protocols analysis have the
property that it is possible to compute, given a positive strand, a
finite set of context pairs that summarizes all possible equalities in
the sense of the next definition. Let us first introduce a notation:
Given a positive strand $s$ we let $P_s$ be the set of context pairs
$(C_1,C_2)$ such that $C_1\cdot s= C_2\cdot s$.

\begin{definition}{\label{def:acceptable}}
  A deduction system \call{D} has the \emph{finite basis property} if
  for each positive strand $s$ one can compute a finite set $P_s^f$ of
  pairs of \call{D}-contexts such that, for each positive strand $s'$:
  $$
  P_s \subseteq P_{s'} \mbox{~~iff~~} P_s^f \subseteq P_{s'}
  $$
\end{definition}
Let us now assume that a deduction system \call{D} has the finite
basis property. There thus exists an algorithm $\call{A}'_\call{D}(s)$
that takes  a positive strand $s$ as input, computes a finite set
$P_s^f$ of context pairs $(C[x_1,\ldots,x_n],C'[x_1,\ldots,x_n])$ and 
returns as a result the $\call{E}$-unification system  $S_s:$ $\set{
  C[x_1,\ldots,x_n] \unif C'[x_1,\ldots,x_n] ~|~(C,C')\in  P_s^f }$. 
For any positive strand $s'=(!m_1,\ldots,!m_n)$ of length
$n$, let $\sigma_{s'}$ be the substitution $\set{x_i\mapsto m_i}_{1\le
  i \le n}$. By definition of $S_s$ we have that $\sigma_{s'}\models
S_s$ if and only if $s'$ is a refinement of $s$. Given the preceding
definition of $\call{A}_\call{D}(s,t)$, we are now ready to present
our algorithm for the compilation of role specifications into active
frames.

\paragraph{Algorithm} Let $r$ be a role specification with
$\strand{r}=(\frac !? M_1,\ldots,\frac !? M_n)$ and let $s=
(!M_1,\ldots,!M_n)$. Let us introduce two notations to simplify the
writing of the algorithm, \textit{i.e.} we write $r(i)$ to denote the
$i$-th labelled message $\frac !?M_i$ in $r$, and $s^i$ to denote the
prefix $(!M_1,\ldots,!M_i)$ of $s$. Compute, for $1\le i \le n$:
$$
T_i=\left\lbrace
  \begin{array}[c]{ll}
    !v_i \text{ with } v_i\unif\call{A}_\call{D}(s^{i-1},M_i) & 
    \text{If }r(i)=!M_i\\
    ?v_i \text{ with } \call{A}_{\call{D}}'(s^i)& 
    \text{If }r(i)=?M_i\\
  \end{array}
\right.
$$
and return the active frame $\varphi_r=(T_i)_{1\le i\le n}$. By
construction we have the following theorem.
\begin{theorem}{\label{th:main}}
  Let \call{D} be a deduction system such that \call{D}-ground
  reachability is decidable and \call{D} has the finite basis
  property.  Then for any executable role specification $r$ one can
  compute a prudent implementation $\varphi$.
\end{theorem}

\section{Examples and Applications}
\label{sec:appli}

Many theories that are relevant to cryptographic protocol design 
satisfy the hypothesis of Theorem~\ref{th:main}. For instance 
let us introduce  the  \emph{convergent subterm theory}:  

\begin{definition}{\label{def:subterm}}
  An equational theory is \emph{convergent subterm} if it admits a
  presentation by a set of equations $\call{E}=\bigcup_{i=1}^n
  \set{l_i =r_i }$ such that $\call{E}$ is a convergent set of rules
  such that each $r_i$ is either a proper subterm of $l_i$ or a ground
  term.
\end{definition}

It is known  (see e.g.~\cite{siva}) that reachability is decidable for  subterm convergent theories. 
It was proved in~\cite{AbadiC06} that any  subterm convergent theory
has the finite basis property too. This is a consequence of Proposition 11 
in~\cite{AbadiC06} that is used by the authors to decide the so-called
static equivalence property for this class of theories. 
We give more details in the Appendix.

Many interesting theories are subterm convergent.  For instance
consider the Dolev-Yao equational theory:
$$
\call{E}_{DY}~\left\lbrace
  \begin{array}[c]{rclr}
    \piun{\paire{x}{y}} & = & x & (P_1) \\
    \pideux{\paire{x}{y}} & = & y & (P_2)\\
    \pdec{\penc{x}{y}}{y^{-1}} & = & x & (D)\\ 
    \symtest{\penc{x}{y}}{y} & = & \mbox{\sc true}& (T_e)\\
    \pairtest{{\paire{x}{y}}}  & = & \mbox{\sc true}& (T_p)\\
  \end{array}
\right.
$$
where $(P_1)$ (resp. $(P_2)$) models the projections on the arguments
of a pair, $(D)$ models the decryption using the inverse key and
$(T_e)$ (resp. $(T_p)$) models that anyone knowing a public key can
test whether a message is encrypted with this key (resp. that anyone
can test whether a message is a pair.)  As a consequence of our
Theorem~\ref{th:main}, for every protocol expressed with functions
satisfying this theory we can compute a prudent implementation.

The equational theory of the \emph{eXclusive-OR} operator $\cdot\oplus\cdot$ is
given by the following set of equations $\call{E}_\oplus$ where $0$ is a
constant and $\oplus,0$ are public functions:
$$
\begin{array}{c}
  \call{E}_\oplus~\left\lbrace
    \begin{array}[c]{rcl}
      (x \oplus y ) \oplus z & = & x \oplus (y \oplus z) \\
      x \oplus y & = & y \oplus x \\
      0 \oplus x  & = & x \\
      x \oplus x  & = & 0\\
    \end{array}
  \right.
\end{array}
$$
This example can be generalized to monoidal theories as follows.  Assume that
all symbols are public and that the signature of a deduction system is equal to
\call{F} $=$\set{+,0,h_1,\ldots,h_n} or \set{+,-,0,h_1,\ldots,h_n} where $+$ is
a binary associative-commutative symbol, $0$ is the identity for $+$, the symbol
$-$ is unary and satisfies the equation $x+ (-x)=0$ and $h_1,\ldots,h_n$ (for
$n\ge 0$) are unary commuting homomorphism on $+$ (\textit{i.e.} such that
$h_i(x+y)=h_i(x)+h_i(y)$ and $h_i(h_j(x))=h_j(h_i(x))$ for $1\le i,j\le n$). Let
us add to the signature $a_1,\ldots,a_k$ the constants appearing in the protocol
narration.

Reachability for this resulting deduction system is decidable (see
e.g. ~\cite{ChevalierKRT03-Lics,cortierlpar}).  We can also show that the deduction system
has the finite basis property: Each ground term in the narration can be
interpreted as an element of the module $(Z[X_1,\ldots,X_n])^k$ as follows:
\begin{itemize}
\item $\lmodel{}a_i\rmodel{}$ is the vector in which only the $i$-th coordinate
  is non-null, and is equal to $1$;
\item $\lmodel{}h_i(t)\rmodel{}=X_i\cdot \lmodel{}t\rmodel{}$, and
  $\lmodel{}t_1+t_2\rmodel{}=\lmodel{}t_1\rmodel{}+\lmodel{}t_2\rmodel{}$, and
  $\lmodel{}-t\rmodel{}=-\lmodel{}t\rmodel{}$.
\end{itemize}
It is routine to check that under these assumptions, we have that:
\begin{itemize}
\item[1.] A context with $m$ holes is interpreted as a linear form mapping
  $((Z[X_1,\ldots,X_n])^k)^m$ to $(Z[X_1,\ldots,X_n])^k$ and with coefficients
  in $Z[X_1,\ldots,X_n]$. These polynomials have positive coefficients whenever
  $-$ is not a public symbol;
\item[2.] In any case we note that any linear form with coefficients in
  $Z[X_1,\ldots,X_n]$ can be written as the difference of two linear forms with
  positive coefficients.
\end{itemize}
Under this interpretation for a positive strand $s$ of length $n$ interpreted as
a vector in $(Z[X_1,\ldots,X_n]^k)^n$ and a pair of contexts $C_1,C_2$ we have
$\lmodel{}C_1\cdot s=C_2\cdot s\rmodel{}$ iff
$(\lmodel{}C_1\rmodel{}-\lmodel{}C_2\rmodel{})(\lmodel{}s\rmodel{})=0$,
\textit{i.e.} there is a mapping from $P_s$ to the set $s^*$ of linear forms $f$
such that $f(\lmodel{}s\rmodel{})=0$. The second remark above shows that this
mapping is surjective. Since $s^*$ is the first syzygy module ~\cite{miller96}
of a linear equation, $s^*$ is also isomorphic to a submodule of
$(Z[X_1,\ldots,X_n]^k)^n$.  Since $Z[X_1,\ldots,X_n]$ is noetherian this syzygy
submodule has a finite generating set $b_1,\ldots,b_l$ that can be computed by
an analogous of Buchberger's algorithm~\cite{miller96}.

Given another strand $s'$ of the same length, if $\set{b_1,\ldots,b_l}
\subseteq (s')^*$ we have also $s^*\subseteq (s')^*$. In other words
we have that $s'$ refines $s$. We thus obtain an algorithm to compute
$P_s^f$ for any strand $s$ that consists in computing a generating set
$b_1,\ldots,b_l$ of $s^*$, write each $b_i$ as the difference
$b_i^+-b_i^-$ of two linear forms with positive coefficients, and
output a set of $n$ pairs of contexts $(C^+_i,C^-_i)$ with
$\lmodel{}C_i^+\rmodel{}=b_i^+$ and $\lmodel{}C_i^-\rmodel{}=b_i^-$
for $1\le i\le n$.

\section{Conclusion}

We have shown how to link the process of compiling protocols 
to excutable roles with formal decision problems. 
This allows us to extend many known results on compilation to the case 
of protocols that are based on more complex cryptographic primitives, 
admitting algebraic properties that are beyond the usual Dolev Yao ones. 

Moreover if the set of symbols occuring in the
protocol can be divided so that each part satisfies an equational
theory with decidable \call{D}-reachability and \call{D} has
the finite basis property, then we can exploit the combination results
from \cite{ChevalierR05,arnaud07} to derive the same properties for 
the union of theories. Therefore the protocol can be prudently
compiled in this case too.

\section*{Appendix}

This Appendix has been added to ease the review. 
We recall some notions and results from ~\cite{AbadiC06} 
and explain why they show that any subterm convergent theory  has 
the finite basis property. 

Let $E$ be a subterm convergent theory. 
The constant $c_E$  introduced in~\cite{AbadiC06} depends only from the equational theory $E$ 
but its exact value is not important for our discussion. 
The size of a term $t$ is the number of vertices in its DAG representation. It is denoted by $|t|$.
To any positive strand $s=(!M_1,\ldots,!M_n)$ 
we can associate a frame with an empty set of free names
in the sense of~\cite{AbadiC06}. This frame 
is $\{M_1/x_1, \ldots , M_n/x_n\}$ and 
will be denoted $s$ too (assuming some variable enumeration).  
We will reformulate or simplify  the results 
from~\cite{AbadiC06}  by  taking into account 
the fact that there are no nonces in the frames in our case. 
Note that $s$ can also be viewed as a substitution. 

Let $st(s)$ be the set of subterms of $s$. 
The set $sat(s)$ (see Definition 3~\cite{AbadiC06}) is  the minimal set 
such that 
\begin{enumerate}
\item $M_1, \ldots, M_n \in sat(s)$;
\item if $N_1, \ldots, N_k \in sat(s)$ and $f(N_1, \ldots, N_k)$ is a subterm of $s$ then 
$f(N_1, \ldots, N_k)\in sat(s)$;
\item if $N_1, \ldots, N_k \in sat(s)$ and $C[N_1, \ldots, N_k] \rightarrow M$ 
where $C$ is a context, $|C| \leq c_E$ and $M$ in $st(s)$ then  $M\in sat(s)$, 
\end{enumerate}

Also Proposition 9 from~\cite{AbadiC06} shows that for every $M\in sat(s)$ 
there exists a term $\zeta_M$ such that $|\zeta_M|_{DAG} \leq  c_E \cdot  |s|$ 
and $\zeta_M\cdot s =_E M$. 

We can now restate Definition 4 from~\cite{AbadiC06} in our framework:

\setcounter{definition}{3}
\begin{definition}
The set $Eq(s)$ is the set of couples:
\[  (C_1[\zeta_{M_1},\ldots,\zeta_{M_k}], C_2[\zeta_{M'_1},\ldots,\zeta_{M'_l}]) \]
such that $(C_1[\zeta_{M_1},\ldots,\zeta_{M_k}]=_E C_2[\zeta_{M'_1},\ldots,\zeta_{M'_l}])\cdot s$,  
$|C_1|,|C_2|\leq c_E$ and the terms $M_i, M'_i$ are in $sat(s)$. 
\end{definition}

Since there are no nonces the set $Eq(s)$ is finite (up to variable renamings).

We recall Lemma 6 and 7 from ~\cite{AbadiC06} with our notations:
\newtheorem{lemma}{Lemma}
\setcounter{lemma}{7}
\begin{lemma}
Let $s,s'$ be two positive strands such that $Eq(s)\subseteq P_s'$. Then for all contexts 
$C_1,C_2$ and  for all terms $M_i,M'_i \in sat(s)$ if 
$C_1[M_1,\ldots,M_k]=  C_2[M'_1,\ldots,M'_l]$ then 
$C_1[\zeta_{M_1},\ldots,\zeta_{M_k}]\cdot s' =_E C_2[\zeta_{M'_1},\ldots,\zeta_{M'_l}] \cdot s'$. 
\end{lemma}

\begin{lemma}
Let $s$ be a positive strand. For every context $C_1$, for every $M_i\in sat(s)$ 
for every term $T$ such that $C_1[M_1,\ldots,M_k]\rightarrow^*_E T$ there is a context 
$C_2$ and terms $M'_i \in sat(s)$ such that $T=C_2[M'_1,\ldots,M'_l]$ 
and for every positive strand $s'$ such that $Eq(s)\subseteq P_s'$ we have 
$C_1[\zeta_{M_1},\ldots,\zeta_{M_k}]\cdot s' =_E C_2[\zeta_{M'_1},\ldots,\zeta_{M'_l}] \cdot s'$.
\end{lemma}

Now we can now extract from Proposition 11 in ~\cite{AbadiC06} the part of the proof  that 
shows our claim:

Assume that $s,s'$ are two positive strands and $Eq(s)\subseteq P_s'$. 
Assume that we have an equality: $M \cdot s =_E N \cdot s$. 
Let $T$ be the common  normal form of $M \cdot s$ and $N \cdot s$ for  the rewrite relation $\rightarrow_E$. 

By Lemma 7 there exists  $M_i\in sat(s)$ and $C_M$ such that \\
$T= C_M[M_1,\ldots,M_k]$ and $M\cdot s' =_E C_M[\zeta_{M_1},\ldots,\zeta_{M_k}]\cdot s'$.\\
By the same lemma there exists $M'_i\in sat(s)$ and $C_N$ such that \\
$T= C_N[M'_1,\ldots,M'_l]$ and $N\cdot s' =_E C_N[\zeta_{M'_1},\ldots,\zeta_{M'_l}]\cdot s'$.\\

Since $C_M[M_1,\ldots,M_k]=C_N[M'_1,\ldots,M'_l]$ we derive from Lemma 6 that:
 \[C_M[\zeta_{M_1},\ldots,\zeta_{M_k}]\cdot s' =_E C_N[\zeta_{M'_1},\ldots,\zeta_{M'_l}] \cdot s'\]

As a consequence we have $M\cdot s'=N\cdot s'$. We can conclude that $P_s \subseteq P_s'$. 
and that the deduction system has the finite basis property by defining  
for all $s$,  $P^f_s$ to be $ Eq(s)$. 

\end{document}